\documentstyle[preprint,prl,aps]{revtex}
\begin{document}
\draft

\tighten

\title{Absolute measurements of the high-frequency magnetic dynamics 
in high-$T_c$ superconductors}
\author{S.M.~Hayden$^{\rm a}$, G.~Aeppli$^{\rm b}$, P.~Dai$^{\rm c}$, 
H.A.~Mook$^{\rm c}$, T.G.~Perring$^{\rm d}$, S.-W.~Cheong$^{\rm e}$, 
Z.~Fisk$^{\rm f}$, F.~Do\u{g}an$^{\rm g}$, T.E.~Mason$^{\rm h}$}
\address{
$^{\rm a}$H.H. Wills Physics Laboratory, University of Bristol, Bristol BS8 1TL, UK \\
$^{\rm b}$NEC Research Institute, Princeton, NJ 08540, USA \\
$^{\rm c}$Oak Ridge National Laboratory, Oak Ridge, TN 37831, USA \\
$^{\rm d}$Rutherford Appleton Laboratory, Chilton, Didcot, OX11 0QX, UK \\
$^{\rm e}$Bell Laboratories, Lucent Technologies, NJ 07974, USA \\
$^{\rm f}$Department of Physics, Florida State University, 
Tallahassee, Florida 32306, USA \\
$^{\rm g}$Department of Material Science and Engineering, 
University of Washington, Seattle, Washington 98195, USA \\
$^{\rm h}$Department of Physics, University of Toronto, Toronto, 
Canada M5S 1A7 \\}

\date{submitted 27 August 1997, version 5 Jan 1998}

\maketitle

\begin{abstract}
We review recent measurements of the high-frequency dynamic magnetic 
susceptibility in the high-$T_c$ superconducting systems 
La$_{2-x}$Sr$_{x}$CuO$_4$ and YBa$_2$Cu$_3$O$_{6+x}$.  Experiments were 
performed using the chopper spectrometers HET and MARI at the ISIS 
spallation source.  We have placed our measurements on an absolute intensity
scale, 
this allows systematic trends to be seen and comparisons with theory to be 
made.  We find that the insulating $S=\frac{1}{2}$ antiferromagnetic parent 
compounds show a dramatic renormalization of the spin wave intensity.  The 
effect of doping on the response is to cause broadenings in wave vector and 
large redistributions of spectral weight in the frequency spectrum.
\end{abstract}

\narrowtext

\section{Introduction}
The collective excitations of the spins in the CuO$_2$ planes of the high 
temperature superconductors display a plethora of behavior \cite{YBCOlit}, perhaps as 
exotic as the bulk properties themselves.  A characterization of 
these excitations is motivated by several considerations.  The spin 
excitations, which can be measured by a well understood and easily 
interpretable probe such as neutron scattering, provide a window on the 
electronic correlations in these complex materials.  Indeed, Cooper pairing 
itself is a correlation between electron spins.  A 
further motivation, especially in view of the unconventional 
pairing \cite{Scalapino95} in 
these materials, is that the magnetic excitations may be involved in the 
pairing attraction.

In the present paper we review our recent 
measurements \cite{Hayden91,Hayden96,Hayden96b,Dai97b} of the dynamic response 
over a wide frequency range i.e.  up to energies of order $2J$.  We find 
that in the superconductors there is significant spectral weight above the superconducting 
gap energy. 
For both superconductors studied, the magnetic response 
$\chi^{\prime\prime}(q,\omega)$ is 
actually stronger for frequencies just above $2\Delta$ than in the parent 
insulating antiferromagnet.  We have been careful to convert our 
measurements into absolute units.  This allows a systematic comparison 
between different systems, compositions and theories to be made.

\section{Experimental}
Our experiments were performed using the HET and MARI spectrometers at the 
ISIS pulsed spallation neutron source of the Rutherford Appleton 
Laboratory.  HET and MARI are direct geometry chopper spectrometers.  A 
pulse of neutrons of approximately 1~$\mu$s duration is produced when a 
800~MeV pulsed proton beam hits a tantalum target.  The neutrons are 
monochromated using a Fermi chopper at 10~m from the target and appropriately 
phased to the proton pulse.  Scattered neutrons are detected using 800 
$^3$He detectors 4~m from the sample.

Neutron scattering directly measures the imaginary part of the generalized 
magnetic susceptibility $\chi^{\prime\prime}({\bf Q},\omega)$. The 
scattering cross-section for an isotropic system 
is \cite{White83},
\begin{eqnarray}
\label{NeutronCross}
\frac{d^2\sigma}{d\Omega \: dE_{f}} &=& (\gamma r_{\rm e})^2 \frac{k_f}{k_i}
\left| F({\bf Q})\right|^2 
\left( 
\frac{2/ \pi g^{2} \mu^{2}_{\rm B}}{1-\exp(-\hbar\omega/kT)}
\right)
\chi^{\prime\prime}({\bf Q},\omega),
\end{eqnarray}
where $(\gamma r_{\rm e})^2$=0.2905 barn sr$^{-1}$\ 
$\mu^{-2}_{\rm B}$, ${\bf k}_{i}$ and ${\bf k}_{f}$ are the incident and 
final neutron wavevectors and $|F({\bf Q})|^2$ is the magnetic form factor.  
Absolute unit conventions were performed by measuring the elastic 
incoherent scattering from a vanadium standard \cite{Windsor81} under the 
same experimental conditions and by measuring the low-frequency coherent 
phonon scattering from the sample \cite{Steinvoll84}.  Throughout this 
paper, we label momentum transfers $(Q_x,Q_y,Q_z)$ in units of
\AA$^{-1}$ by their reciprocal space positions {\bf Q}=$(h,k,l)$=
$(2\pi Q_{x}/a,2\pi Q_{y}/b,\pi Q_{z}/c)$.  Following previous practice, we use the 
{\it orthorhombic\/} nomenclature to label reciprocal space in the 
La$_{2-x}$Sr$_{x}$CuO$_4$ system and {\it tetragonal\/} nomenclature in the 
YBa$_2$Cu$_3$O$_{6+x}$ system.  This means that the CuO$_2$ planes are parallel 
to (010) in La$_{2-x}$Sr$_{x}$CuO$_4$ and (001) in YBa$_2$Cu$_3$O$_{6+x}$. 
Details of the samples are given 
elsewhere \cite{Hayden91,Hayden96,Hayden96b,Dai97b}. In the case of 
the La$_{2-x}$Sr$_{x}$CuO$_4$ system, data were collected with the (001) 
plane coincident with the principal scattering plane of the spectrometer.  
For YBa$_2$Cu$_3$O$_{6+x}$, the $(1\overline{1}0)$ plane was used.

\section{The Quantum Antiferromagnets}
We first discuss the parent antiferromagnets 
La$_2$CuO$_4$ and YBa$_2$Cu$_3$O$_{6.15}$.  For the purpose of this paper, we 
model the systems as a set of weakly coupled CuO$_2$ layers or bilayers.  In this 
case, the high-frequency spin excitations can be described by the Heisenberg 
Hamiltonian for a single CuO$_2$ layer or bilayer,
\begin{eqnarray}
\label{Heisenberg}
H & = & \sum_{ij} J_{\parallel} \, {\bf S}_i \cdot {\bf S}_j +
        \sum_{ij^{\prime}} J_{\perp} \, {\bf S}_i \cdot {\bf S}_{j^{\prime}}.
\end{eqnarray}
The first term in Eq.~\ref{Heisenberg} represents the nearest-neighbor 
coupling between Cu spins in 
the same CuO$_2$ plane.  The second term (not present in the single 
layer compound) represents the coupling between nearest-neighbor Cu spins 
in different layers.

In the case of the single layer compound ($J_{\perp}=0$), conventional  
spin-wave theory in the classical large-$S$ limit yields the 
transverse dynamic susceptibility
\begin{eqnarray}
\label{ChiLa}
\chi^{\prime\prime}_{\perp}({\bf Q},\omega) & = & Z_{\chi} \, 
\frac{\pi}{2}\, g^2 \, \mu^2_{\rm B} 
   \, S \,
   \left( \frac{1-\gamma({\bf Q})}{1+\gamma({\bf Q})} \right)^{1/2}
   \delta \left( \hbar\omega \pm \hbar \omega ({\bf Q}) \right),
\end{eqnarray}
where,
\begin{eqnarray}
\label{wLa}
\hbar \omega ({\bf Q}) & = & 2 Z_{c} J_{\parallel} 
  \left[ 1 - \gamma^2({\bf Q}) \right]^{1/2},
\end{eqnarray}
and $\gamma({\bf Q})=\cos(\pi h)\cos(\pi l)$.  We have included a ``quantum 
renormalization'' of the overall amplitude $Z_{\chi}$.  In the 
conventional linear spin-wave model applicable for large $S$, $Z_{\chi}=1$.  
For small $S$, quantum corrections \cite{Singh89,Igarashi92} become important 
(the Neel state is 
not a good approximation to the ground state) leading to a renormalization 
of the overall scale of the spin-wave dispersion and to a reduction in 
magnetic response with respect to the classical spin-wave theory.  The 
renormalization of the overall scale can be included in the exchange constant 
$J^{*}=Z_c J$, where $J$ is the exchange constant occurring in 
Eq.~\ref{Heisenberg}.  The value of 
$Z_{\chi}$ {\it can \/} be obtained by neutron scattering, if measurements 
are placed on an absolute intensity scale.  However, $Z_c$ {\it cannot\/} be measured 
directly from inelastic neutron scattering and must be estimated from theory.  In the 
case of the $S=\frac{1}{2}$ square-lattice antiferromagnet, 
Singh \cite{Singh89} and Igarashi \cite{Igarashi92} have estimated $Z_c=1.18$ 
and $Z_{\chi}$=0.51 based on a $1/S$ expansion.

The presence of the second term (i.e.  $J_{\perp} \neq 0$) in 
Eq.~\ref{Heisenberg} leads to the existence of two branches in the 
spin-wave dispersion which can be labeled according to whether neighboring 
spins in different planes rotate in the same direction (``acoustic'' or 
``odd'' mode) or in opposite directions (``optical'' or ``even'' mode) 
about their time-averaged (ordered) directions.  In the conventional linear spin-wave 
approximation, the acoustic and optic modes have the response 
functions per formula unit\cite{Tranquada92}
\begin{eqnarray}
\label{ChiAc}
\chi^{\prime\prime}_{\rm ac}({\bf Q},\omega) & = 
   & Z_{\chi} \, \pi\, g^2 \, \mu^2_{\rm B} \, S \,
   \left( \frac{1-\gamma({\bf Q})+J_{\perp}/2 J_{\parallel}}{1+\gamma({\bf Q})} 
   \right)^{1/2}  \\
   & & \times 
   \sin^2 \left(\frac{\pi \, \Delta z \, l}{c} \right)  
   \delta \left( \hbar\omega \pm \hbar \omega_{\rm ac} ({\bf Q}) \right)
\end{eqnarray}
and
\begin{eqnarray}
\label{ChiOp}
\chi^{\prime\prime}_{\rm op}({\bf Q},\omega) & = 
   & Z_{\chi} \, \pi\, g^2 \, \mu^2_{\rm B} \, S \,
   \left( \frac{1-\gamma({\bf Q})}{1+\gamma({\bf Q})+J_{\perp}/2J_{\parallel}} 
   \right)^{1/2}  \\
   & & \times
   \cos^2 \left(\frac{\pi \, \Delta z \, l}{c} \right)  
   \delta \left( \hbar\omega \pm \hbar \omega_{\rm op} ({\bf Q}) \right),
\end{eqnarray}
respectively.  The dispersion relations are
\begin{eqnarray}
\hbar \omega_{\shortstack[c]{\rm ac \\ op}} ({\bf Q}) & = &
    2 Z_c J_{\parallel} \left\{ 1-\gamma^2({\bf Q})+J_{\perp}/J_{\parallel}
\left[ 1 \pm \gamma({\bf Q}) \right] \right\}^{1/2},
\end{eqnarray}
where $\gamma({\bf Q})=\frac{1}{2}[\cos(2 \pi h)+\cos(2 \pi k)]$ and 
$\Delta z$= 3.2~\AA\ is the separation of the CuO$_2$ planes in a bilayer.  
The inter-planar coupling term in Eq.~\ref{Heisenberg} leads to no 
additional dispersion along the $z$-direction, only a modulation in the 
amplitude of the response which nevertheless can be used to distinguish 
between the two modes.  
 
\subsection{La$_2$CuO$_4$}
The two-dimensionality of the scattering in La$_2$CuO$_4$ means that we are 
able to cut through the spin waves at several energy transfers for a single 
spectrometer setting.  Fig.~\ref{LaSr300} shows data collected for 
$E_i=$300~meV and ${\bf k}_i \parallel (010)$ on the MARI spectrometer.  
Panels (b)-(f) show constant energy cuts along the (1,0,0) direction.  A 
spin-wave peak is observed near $h=1$.  Twin peaks due to spin waves 
propagating in opposite directions are not observed due to the poor 
out-of-plane resolution in the (001) direction.  The broadening in 
the peak at higher frequencies is due to the spin-wave dispersion.  A 
convenient way to display the data in Fig.~\ref{LaSr300} is as a local- or 
wavevector-integrated susceptibility 
$\chi^{\prime\prime}(\omega)= \int_{\rm BZ} \, 
\chi^{\prime\prime}({\bf Q},\omega) \, d^3Q/ \int d^3Q$, 
this is shown in 
Fig.~\ref{LaSrlocal}(a).  The solid lines in Fig.~\ref{LaSr300}(b)-(f) and 
Fig.~\ref{LaSrlocal} are fits to the spin-wave model described by 
Eq.~\ref{ChiLa}.  Quantitative analysis yields an exchange constant
consistent with our previous 
determination \cite{Hayden91} $J^{*}=156 \pm 5$~meV and $Z_{\chi}=0.39 \pm 0.1$. 
This is in agreement with the calculation of Igarashi \cite{Igarashi92} and 
demonstrates the importance of quantum corrections in this system.

\subsection{YBa$_2$Cu$_3$O$_{6.15}$}
As discussed above, the bilayer nature of this material results in two 
collective modes with additional 
structure to $\chi^{\prime\prime}(\bf Q,\omega)$ as compared with the 
single layer material.  The optic mode has a gap at the 2-D magnetic zone 
centers such as $(\frac{1}{2}\frac{1}{2}l)$ of 
$\hbar \omega_{g} = 2\sqrt{J_{\perp}^{*} J_{\parallel}^{*}}$. Its intensity displays
an overall modulation $\cos^2({\pi \, \Delta z \, l/c})$ which has maxima at 
positions $l=$ $0,3.7,7.3,\ldots$.  Correspondingly, the acoustic 
mode has an overall modulation of $\sin^2({\pi \, \Delta z \, l/c})$ and 
therefore has maxima at $l=$ $1.8,5.5,\ldots$.  We are able to probe the spin
waves at various values of $\hbar \omega$ and $l$ by varying the incident 
energy.  Fig.~\ref{YBCOlow}(b)-(e) show data collected for different energy 
transfers at $l$ values corresponding to positions where the scattering is 
predominately acoustic.  As in La$_2$CuO$_4$, a peak is 
observed near the magnetic zone center, for all energy transfers investigated, 
due to propagating spin waves.  Similar cuts probing $l$ values where the optic 
mode dominates, are shown in Fig.~\ref{YBCOlow}(g)-(j).  
In this case, no peak is observed at the lowest energies because 
these energies are below the optic gap. A more convenient way to look at 
these data 
is in the form of a local susceptibility (see above).  Fig.~\ref{YBCOlocal} 
shows the contributions of the two branches to the local susceptibility 
extracted from the data in Fig.~\ref{YBCOlow}.  When the data are plotted 
in this way, we see that the optic and acoustic contributions are equal at 
higher frequencies.  While below $\hbar \omega_{g}$, the  
contribution from the optical mode is zero within the experimental error. 
A detailed analysis \cite{Hayden96b} yields a value for the optic gap of $\hbar 
\omega_g=74\pm5$~meV.  A similar value for the optic gap was obtained
using reactor-based instrumentation \cite{Reznik96}.

In order to determine the exchange constant $J_{\perp}^{*}$ describing the 
coupling of spins in different planes, we also need to know 
$J_{\parallel}^{*}$ 
because the optic gap is $\hbar \omega_{g} = 2\sqrt{J_{\perp}^{*} 
J_{\parallel}^{*}}$.  The in-plane exchange coupling, $J_{\parallel}^{*}$ can be 
obtained from a measurement of the high-frequency spin waves.  
Fig.~\ref{YBCOhigh} shows data collected with the higher incident energy
$E_i=$600~meV.  We note that for energy transfers above 245~meV there is 
little variation in the intensity, suggesting that this energy is close to 
the zone boundary energy.  A simultaneous resolution-corrected fit of the 
linear spin wave 
model (Eqs.~(\ref{ChiAc})-(\ref{ChiOp})) to all our data yields values for 
the exchange constants of $J^{*}_{\parallel}=$ $125\pm5$~meV and 
$J^{*}_{\perp}=$ $11 \pm 2$~meV.  Further, since our measurements are in 
absolute units we are able to estimate the amplitude renormalization.  We 
find a value of $Z_{\chi}=$ $0.4\pm0.1$, as for La$_2$CuO$_4$, again the quantum renormalization 
is close to the value predicted by a $1/S$ expansion ($Z_{\chi}=0.51$).

\section{The Superconductors}
Of particular interest is the nature of the spin fluctuations 
\cite{Yamada95,Hayden96} for metallic and superconducting compositions .  
We have therefore studied the compounds La$_{1.86}$Sr$_{0.14}$CuO$_4$ (a 
composition close to optimal doping) and YBa$_2$Cu$_3$O$_{6.6}$ (an underdoped 
superconductor).  In both cases we find strong high-frequency spin 
fluctuations in the superconducting state.

\subsection{La$_{1.86}$Sr$_{0.14}$CuO$_4$}
Fig.~\ref{LaSr300}(h)-(l) shows data collected for La$_{1.86}$Sr$_{0.14}$CuO$_4$
under the same conditions as Fig.~\ref{LaSr300} and in the {\it same 
units\/}.  Doping has a dramatic effect on the magnetic excitations: at low 
frequencies, the peaks are broader in reciprocal space and at higher 
frequencies there is a large suppression of the intensity.  At the 
highest frequencies, the peaks appear to be disappearing more rapidly in 
the superconductor than the insulator.  This can be seen clearly when we 
extract and plot the local susceptibility in Fig.~\ref{LaSrlocal}(b).  When we 
include low frequency data from a reactor-based experiment, it becomes 
clear that the local response is peaked near 22~meV.  Thus doping leads to 
a shift of spectral weight to intermediate energies. 

\subsection{YBa$_2$Cu$_3$O$_{6.6}$}
The low frequency dynamics of superconducting YBa$_2$Cu$_3$O$_{6+x}$ have been 
the subject of considerable investigation \cite{YBCOlit}.  The 
YBa$_2$Cu$_3$O$_{6.6}$($T_c=62.7$~K) sample used 
in the current study is of high quality: it shows a sharp ``resonance 
peak'' \cite{Dai96} at 34~meV on entering the superconducting state and has 
recently been shown \cite{Dai97} to develop incommensurate peaks in 
$\chi^{\prime\prime}({\bf q},\omega)$ for frequencies near 25~meV.  We concentrate here
on the nature of the dynamics for temperatures just above the 
superconducting transition temperature.  Fig. ~\ref{YBCO6.6low} shows 
inelastic data collected for $T=80$~K on the MARI spectrometer in a similar
manner to that in Fig.~\ref{YBCOlow}.  The reader should note that the energies 
chosen for Fig.~\ref{YBCO6.6low} are different from those chosen for 
Fig.~\ref{YBCOlow} 
because of the different detector positions on the HET and MARI 
spectrometers. By assuming that the scattering has a Gaussian wavevector 
dependence for each energy, we have extracted the frequency variation of 
the acoustic and optical contributions to the local susceptibility 
$\chi^{\prime\prime}(\omega)$ from data such as those shown in the figure.  
Fig.~\ref{YBCO6.6local} shows the results.

Comparison of Figs.~\ref{YBCO6.6low} and \ref{YBCO6.6local} with 
Figs.~\ref{YBCOlow} and \ref{YBCOlocal} shows that the magnetic response
has changed dramatically between the metal and the insulator.  This is well-known at low frequencies \cite{YBCOlit}, as in  La$_{2-x}$Sr$_{x}$CuO$_4$, there
is a large broadening of the low frequency peak in wavevector.  More 
dramatic is the redistribution of spectral weight in energy revealed by the
local susceptibility in Fig.~\ref{YBCO6.6local}.  Inspection of 
Fig.~\ref{YBCOlocal} and Fig.~\ref{YBCO6.6local}, which are in {\it the same units}
and therefore can be directly compared, shows that doping causes a shift of
spectral weight from high frequencies (above 100~meV) into the range around
50-100~meV. The gap in the optic fluctuations remains and is possibly 
slightly reduced.  Reactor-based measurements yield a similar value for the
optic gap\cite{Bourges97}. 

\section{Discussion}
Our measurements of $\chi^{\prime\prime}(\omega)$ on La$_2$CuO$_4$ and 
YBa$_2$Cu$_3$O$_{6.15}$ are summarized in Fig.~\ref{LaSrlocal}(a) and 
Fig.~\ref{YBCOlocal}.  We note that observed $\chi^{\prime\prime}(\omega)$ for
La$_2$CuO$_4$ and the acoustic branch of
YBa$_2$Cu$_3$O$_{6.15}$ are of approximately equal intensity: the units of the susceptibility are expressed per
formula unit in both cases, YBa$_2$Cu$_3$O$_{6+x}$ has two Cu in the CuO$_2$ 
planes per formula unit.
When we compare our measurements on the quantum 
antiferromagnets La$_2$CuO$_4$ and YBa$_2$Cu$_3$O$_{6.15}$,
we find that, in both cases, linear spin-wave theory accounts well for the 
dispersion and amplitude variation.  However, an overall renormalization of 
the amplitude $Z_{\chi}$ must be included to account for the observed 
intensity of the one-magnon response.  Our observations agree with calculations 
of $Z_{\chi}$=0.51 based on a $1/S$ expansion. Although we are 
able to explain the one-magnon excitation spectrum, the $1/S$ expansion also
predicts additional spectral weight \cite{Igarashi92} which has not yet been observed.

We now turn the metals La$_{1.86}$Sr$_{0.14}$CuO$_{4}$ and 
YBa$_2$Cu$_3$O$_{6.6}$. The response of the copper spins obeys the
sum rule,
\begin{eqnarray}
\label{sumrule}
\left<m^2\right> & = & \frac{3 \hbar}{\pi} 
   \int_{-\infty}^{\infty} \, 
   \frac{\chi^{\prime\prime}(\omega) \, d\omega}
        {1-\exp(-\hbar\omega/kT)}
\end{eqnarray}  
where $\left<m^2\right>$ is the mean squared moment on the Cu sites.  
In both superconductors studied here, doping and the consequential
metal-insulator transition led to a large 
redistribution of spectral weight and formation of a peak in 
$\chi^{\prime\prime}(\omega)$ at intermediate frequency.  In the light
of the sum rule, the existence of the peak in itself is not unexpected since 
spectral weight is lost from the Bragg peak present in the insulator and 
the high frequency response is strongly suppressed.  What is surprising is the relatively small energy
scale over which the response is distributed compared with other paramagnetic
metals.  Such small electronic energy scales are, of course, characteristic of the
cuprate superconductors.

Work at ORNL is supported by US-DOE under Contract No. DE-AC05-96OR22464
with Lockheed Martin Research, Inc. We are grateful for the finacial 
support of the UK-EPSRC and NATO.

\begin{figure}
\caption{(a)-(f) Magnetic scattering from La$_2$CuO$_4$ and (g)-(l) from 
La$_{1.86}$Sr$_{0.14}$CuO$_4$ \protect\cite{Hayden96}.  
All scans are in the same absolute units. 
Note the magnetic peak is broader in the superconductor and suppressed at
higher frequencies.}
\label{LaSr300}
\end{figure}

\begin{figure}
\caption{Local susceptibility \protect\cite{Hayden96} derived from data such as that in 
Fig.~\protect\ref{LaSr300} (closed circles) and from reactor-based 
measurements \protect\cite{Aeppli97}.  Note huge redistribution of spectral
weight caused by doping.}
\label{LaSrlocal}
\end{figure}

\begin{figure}
\caption{Constant energy scans showing magnetic scattering from 
YBa$_2$Cu$_3$O$_{6.15}$ for wavevectors ${\bf Q}=(h,h,l)$.  (b)-(e) $l$ 
chosen to emphasize acoustic modes. (g)-(j) $l$ chosen to emphasize 
optic modes.  The onset of scattering at the optic position is about 
74~meV.}
\label{YBCOlow}
\end{figure}

\begin{figure}
\caption{Local or 
wavevector-integrated energy-dependent magnetic susceptibility at
acoustic and optic $l$-positions in YBa$_2$Cu$_3$O$_{6.15}$. No integration
over $l$ has been performed. Points have
been obtained by integrating over the spin-wave peaks and correcting for 
the Cu$^{2+}$ magnetic form factor, Bose factor and instrumental resolution.
Solid lines are a fit to the spin wave model described in the text. }
\label{YBCOlocal}
\end{figure}

\begin{figure}
\caption{(a)  Dispersion relation of YBa$_2$Cu$_3$O$_{6.15}$.  Closed 
circles and solid line are acoustic mode.  Open circles and dashed line are
optic mode.  (b)-(e) Constant energy scans showing the high-frequency magnetic 
scattering from YBa$_2$Cu$_3$O$_{6.15}$. }
\label{YBCOhigh}
\end{figure}

\begin{figure}
\caption{Constant energy scans showing magnetic scattering from 
YBa$_2$Cu$_3$O$_{6.6}$ for wavevectors ${\bf Q}=(h,h,l)$.  (a)-(c) $l$ 
chosen to emphasize acoustic modes. (d)-(f) $l$ chosen to emphasize 
optic modes.}  
\label{YBCO6.6low}
\end{figure}

\begin{figure}
\caption{Acoustic and optic contributions to the local susceptibility 
(see Fig.~\protect\ref{YBCOlocal}) in 
YBa$_2$Cu$_3$O$_{6.6}$. Solid lines are a guide to the eye.}
\label{YBCO6.6local}
\end{figure}

\begin{table}
\caption{The table summarizes experimentally-determined spin-wave parameters for 
La$_{2-x}$Sr$_{x}$CuO$_4$ and YBa$_2$Cu$_3$O$_{6+x}$. In the case of the 
superconductors, $J_{\parallel}^{*}$ and $Z_c$ are determined from the highest 
frequencies studied.}
\begin{tabular}{lcccc} 
& $J_{\parallel}^{*}$ & $J_{\perp}^{*}$  &$Z_{\chi}$ & $\hbar \omega_{g}$ \\
\hline
La$_2$CuO$_4$   & $156\pm 5$ meV       &           & $0.39\pm0.1$  &  \\
La$_{1.86}$Sr$_{0.14}$CuO$_4$ & $130\pm 5$ meV &   & $0.15\pm0.06$ &  \\ 
YBa$_2$Cu$_3$O$_{6.15}$ & $125\pm5$~meV & $11\pm2$meV & $0.4\pm0.1$ & 
       $74\pm5$~meV \\
YBa$_2$Cu$_3$O$_{6.6}$&$125\pm20$meV  & &$0.24\pm0.12$ & $\sim$ 60~meV \\   
\end{tabular}
\end{table}

\end{document}